# A new conversational interaction concept for document creation and editing on mobile devices for visually impaired users

Alireza Darvishy[1], Hans-Peter Hutter[1], Edin Beljulji[1], Zeno Heeb[1]

[1]Zurich University of Applied Sciences, 8400 Winterthur, Switzerland

**ABSTRACT**

This paper describes the ongoing development of a conversational interaction concept that allows visually impaired users to easily create and edit text documents on mobile devices using mainly voice input. In order to verify the concept, a prototype app was developed and tested for both iOS and Android systems, based on the natural-language understanding (NLU) platform Google Dialogflow. The app and interaction concept were repeatedly tested by users with and without visual impairments. Based on their feedback, the concept was continuously refined, adapted and improved on both mobile platforms. In an iterative user-centred design approach, the following research questions were investigated: Can a visually impaired user rely mainly on speech commands to efficiently create and edit a document on mobile devices? User testing found that an interaction concept based on conversational speech commands was easy and intuitive for visually impaired users. However, it was also found that relying on speech commands alone created its own obstacles, and that a combination of gestures and voice interaction would be more robust. Future research and more extensive useability tests should be carried out among visually impaired users in order to optimize the interaction concept.

**Keywords:** Visual impairment, mobile devices, non-visual interaction, NLP, speech input, speech output, document creation

## INTRODUCTION AND RELATED WORKS

According to WHO, there are more than 250 million people living with a visual impairment worldwide (WHO, 2012) Many people with visual impairments use assistive technologies such as screen readers to read digital content aloud. A screen reader is a software application that allows individuals who are blind or have low vision to access and use the features and functions of a computer or smartphone. Screen readers work by providing synthesized speech output of the text and other information displayed on the screen, as well as allowing the user to interact with the device using keyboard commands or gestures. There are different screen readers available for mobile devices, such as "Voice Over" in iOS and "TalkBack" in Android.

In addition to speech output, recent advances in speech technologies have also enabled users to harness speech input to interact with their mobile devices. For example, users are able to control their mobile devices solely with voice commands, e.g. using "Voice Control" in iOS or "Voice Access" in Android devices. As a use case, users can use voice commands to edit a text on their mobile devices. These are useful tools for many people with visual impairments, when compared to other text entry methods such as onscreen keyboards, gesture-based text entry (wherein the user draws the desired letter/character on the screen using a finger or stylus), or braille-based digital keyboards (e.g. Mattheiss et. al, 2015; Oliveria et. al, 2011). However, despite the usefulness of this new software, current mainstream speech input programs are not designed with visually impaired users in mind. As a result, many accessibility issues remain. For example: In iOS, the Voice Control command needed to delete a certain section of text is the phrase "delete that" – a command which assumes that the user can see which section of text is currently highlighted.

Speech input has been found to be one of the most efficient text entry methods for users with visual impairments. A 2013 user study by Azenkot and Lee found that speech input was almost five times faster for users with visual impairments than using a standard touch keyboard. However, the same study noted that, despite the relative ease of entering text, errors in speech recognition remained a major problem: on average, when composing a text, participants spent 80% of their time finding and correcting errors (Azenkot & Lee, 2013).

Text editing and error correction can present significant barriers to visually impaired users, particularly with regards to unstructured and inaccessible document formats. Structural elements such as headings, tables, and lists, are not only more difficult to create (requiring additional commands beyond simple dictation), but they are also important elements in document accessibility, necessary for efficient navigation via screen readers (see e.g. Rajkumar et al. 2020, Darvishy et al. 2011, Munteanu et al. 1995).

In late 2015, Google rolled out a feature called "Voice Typing", a speech-to-text dictation feature that can be used in conjunction with Google Docs on the Chrome browser (Moynihan, 2016). This technology could hold great promise for visually impaired users because, unlike many other dictation programs and apps, it includes features such as inserting structural elements (headings, tables, lists, etc.), jumping to a desired word/sentence in the document, and adjusting formatting using speech commands. However, Google Docs creation using Voice Typing is currently only available for desktop use via Chrome. It is also not designed with visually impaired users in mind – for example, users must click the microphone icon to start dictation.

## CONCEPT DEVELOPMENT

The major goal of this innovative interaction concept is to allow users to easily create documents in a conversational manner via speech input, including commonly used formatting features such as headings and subheadings, indentation, bullet lists, enumerations, etc. The interaction concept also allows users to edit the document in a conversational manner, e.g. adding, removing or changing text by directly referring to them. Consequently, speech output was also implemented to help users navigate and orient themselves within the document without relying on a cursor or other visual cues.

In order to develop the conversational interaction model for that in a user-centered manner, a first user research and analysis of user needs was carried out. Subsequently, based on the user needs identified, a first design and prototype iteration was implemented. With this prototype, extensive user tests were carried out.

In order to define the user needs, initial informal interviews were conducted with visually impaired potential users. These discussions revealed that the major tasks needed for mobile document creation, such as text dictation, editing, and basic structuring, are particularly cumbersome and time-consuming for visually impaired users. Based on this feedback, an initial concept was developed by defining context scenarios and sample dialogues for a prototype app called "Spectra".

### Context Scenarios

A series of context scenarios were defined to demonstrate some envisioned interactions with the prototype app. In each context scenario, fictional personas were described performing a task using the app. For example, one such context scenario, "proofreading a document," reads as follows:

Angelica works as a technical assistant at a university. It's Monday afternoon and she has been asked to proofread a few documents for technical accuracy before they are submitted next week. As Angelica has an optic nerve disease, she cannot do this visually. Instead, she launches the Spectra app on her smartphone, opening her first document for today. She can sit back and relax in her chair while the app reads her a brief overview of the headings in this document. Since Angelica is only interested in the technically important chapters, she instructs the app to jump directly to a specific chapter and read it aloud. With the stop command, she interrupts the reading, "Please repeat the last sentence". Spectra reads her the last sentence again. Because the content of the sentence is incorrect, she uses a voice command to insert a comment into the document: "Insert comment: This sentence should be reformulated". Spectra replies: "Comment inserted: This sentence should be reformulated". After Angelica's reply, "go on," Spectra continues with reading on.

The following context scenarios were considered:
- Create a shopping list
- Write a letter
- Write a report
- Review and edit a document

## Sample dialogues

A number of sample dialogues were created to conceptualize concrete interactions between the user and the system, showing both the user's verbal input and the system's speech output. By default, the app is programmed to confirm/read back each command or dictated sentence, so that the user can detect and correct any errors or misunderstandings. A snippet of such a sample dialogue is shown in Table 1.

Table 1. Snippet of sample dialogue, in which the user writes and edits a document about a new product (an anti-theft system)

| Turn | Utterance |
|---|---|
| user | Title «anti theft system» |
| system | Document title "Anti Theft System" |
| user | Heading one «introduction» |
| system | Heading 1 «instructions» |
| user | Replace «instructions» with «instruction» |
| system | Heading 1 «Introduction» |
| user | Dictation mode |
| system | Dictation mode started |
| user | This new system should achieve protection against burglary comma both in the absence and presence of residents period |
| system | This new system should achieve protection against burglary, both in the absence and presence of residents. |
| user | Insert "control" before "system" |
| system | This new control system should achieve protection against burglary, both in the absence and presence of residents. |

## Interaction Concept

The interaction concept envisages two modes: the dictation mode and the command mode. In the dictation mode, the running text of the document is dictated by voice. Hence, this mode mainly provides speech-to-text (STT) where the voice input is interpreted as text. The text is entered utterance by utterance. The dictation mode also allows punctuation information within the dictated text, e.g., commas, semicolon, etc (see Table 1). The system automatically reads back the recognised utterance as confirmation, including formatting and punctuation information

(except normal periods and question marks at the end of a sentence, which are indicated by prosodic means)

After readback of the recognised utterance the command mode becomes automatically active where the user can immediately edit the text just entered by replacing, moving, inserting or deleting words. The user can thereby directly refer to the words entered before, e.g. "Insert 'control' before 'system' in Table 1. If the user just goes on with dictating, the system automatically switches back to dictation mode. The user can explicitly switch between dictation and command mode with the corresponding commands.

With the command mode the dictated text can be reviewed, edited, formatted and structured (headings, paragraphs, lists etc.) Erroneous entries or commands can be easily undone with the corresponding commands. There are also various jump commands to navigate quickly in the document. To get a better overview of the content of a document, users can also input various read commands, i.e. to request that certain words, sentences, paragraphs or the entire document be read aloud. Commands related to a relative position, e.g., "delete last word" are interpreted with respect to the current focus (virtual cursor). This is set at the end of the last word read back or at the position of the last editing command.

Alternatively to the pure voice interaction, the app also supports screen readers, which can be used to read out elements on the touchscreen using tapping gestures. An export function is also available to share the dictated documents with other people.

## Prototype Implementation

In order to efficiently implement a prototype of the interaction concept, a distributed speech recognition approach is used based on the local STT components on the smartphone (iOS or Android). The recognised text is then sent to the remote intent-based Dialogflow NLU service (Sabharwal & Agrawal, 2020) for the recognition and interpretation of the commands, after having set the right context for the intent-based interpretation of the commands. The recognised commands are finally sent back to the mobile device, where they are executed by the app, or the recognized sentence is read back to the user. Immediately after that, the next recognition loop starts.

## User Testing

To test the usability of the initial prototype, a group of test users was established which included both visually impaired and non-visually impaired users. Each test user was asked to download the Spectra app and was then sent a sample text. The test task was to re-create the sample text using only the app (alongside their device's built-in screen reader, if needed). Test users then submitted the resulting text document to the authors. All test users also submitted unstructured written feedback describing their overall experience with the app and noting any obstacles they encountered. The current iteration of the Spectra app implements changes and improvements based on this feedback.

## RESULTS AND DISCUSSION

This ongoing work has delivered two major findings so far:

1) Using content-based commands, as described below, are most efficient for visually impaired users in many cases, and is often more practical than cursor-based commands.
2) Relying only on speech commands and voice output functions is insufficient for visually impaired users. Rather, a combination of gestures and voice interaction is ideal.

In the planning and the first prototype implementation, the app primarily took "cursor-based" commands, such as "select word" or "delete sentence" which were interpreted relative to the current cursor position. In this case, if a word is to be deleted, the cursor must be first moved to the correct position in the text, as in normal text editors. However, based on user feedback, we have learned that such commands are not suitable for non-visual text editing. In order to find the desired word or sentence, users either had to have large sections of content read aloud and then stop at the right place, or had to use navigation commands to move around. The alternative to cursor-based commands are more "content-based commands", which more closely mimic how people would communicate with one another. For example, the command "Replace x with y" is more reminiscent of conversational phrases like "sorry, I meant to say x, not y". These commands do not require the position to be specified and are less cumbersome than navigating to find "x" and giving the commands "delete word" and "dictation mode: y". In the newest prototype, we have thus implemented content-based commands in which the old and new content is specified directly within the command, so that no navigation commands must be executed first. However, because content-based commands are not always easier to use or can be misinterpreted by the system, standard cursor-based commands have also been kept in the prototype, giving users the choice of using either kind.

The current version of the app includes commands to insert text or structured content such as paragraphs, lists or headings, as well as commands to undo or amend incorrect entries. There are also commands to have parts of the document or the entire document read aloud. The reading can be interrupted using a single-tap gesture in order to make changes at the desired point. Navigation commands such as "start of paragraph", "end of paragraph" etc., are also available to change the current position in the document.

User testing also identified screen reader support, particularly iOS VoiceOver support, as a desirable feature. This support means that changes to the text can also be made using the device's built-in screen reader functions, whenever Spectra and the screen reader are running simultaneously. Although iOS and Android both feature built-in screen readers, the overall accessibility support on iOS is significantly higher and our target group is therefore increasingly using the iOS operating system; as such, the focus for the current iteration of Spectra is on VoiceOver support.

Delegating the NLU to Dialogflow brought several advantages, since the analysis takes place independently of the native mobile system, and without the need for a local agent. As a result, the Dialogflow agent has significantly more resources available than would be possible on a single mobile device. However, of disadvantage is that there is a delay before the data is sent to and evaluated by Dialogflow, and the result is received. While this process is not significantly slower on Android, it takes longer on the iOS operating system. As a result, the speed of use is lower than would be the case with a local agent.

One major research question remains unanswered: What is the optimal combination of speech commands, gestures and screen reader output to allow visually impaired users to efficiently create and edit text documents on mobile devices?

This question will be addressed in the future development of Spectra, and with more extensive user testing.

## **REFERENCES**

1. Azenkot, S. and Lee, N. (2013). Exploring the use of speech input by blind people on mobile devices. Proceedings of the 15th International ACM SIGACCESS Conference on Computers and Accessibility, ASSETS 2013.
2. Azenkot, S. Wobbrock, J.O. Prasain, S. Ladner, R.E. (2012). Input finger detection for nonvisual touch screen text entry in Perkinput. Proceedings of graphics interface 2012, pp. 121-129.
3. Darvishy, A., Hutter, HP., Mannhart, O. (2011). Web Application for Analysis, Manipulation and Generation of Accessible PDF Documents. In: Stephanidis, C. (eds) Universal Access in Human-Computer Interaction. Applications and Services. UAHCI 2011. Lecture Notes in Computer Science, vol 6768. Springer, Berlin, Heidelberg.
4. Mattheiss, E. Regal, G. Schrammel, J. Garschall, M. Tscheligi, M. (2015). EdgeBraille: Braille-based text input for touch devices. Journal of Assistive Technologies.
5. Darvishy, A., Leemann, T., Hutter, HP. (2012). Two Software Plugins for the Creation of Fully Accessible PDF Documents Based on a Flexible Software Architecture. In: Miesenberger, K., Karshmer, A., Penaz, P., Zagler, W. (eds) Computers Helping People with Special Needs. ICCHP 2012. Lecture Notes in Computer Science, vol 7382. Springer, Berlin, Heidelberg.
6. Darvishy, A., Hutter, HP. (2013). Comparison of the Effectiveness of Different Accessibility Plugins Based on Important Accessibility Criteria. In: Stephanidis, C., Antona, M. (eds) Universal Access in Human-Computer Interaction. Applications and Services for Quality of Life. UAHCI 2013. Lecture Notes in Computer Science, vol 8011. Springer, Berlin, Heidelberg.
7. Moynihan, Tim. (2016) "Now You Can Edit Google Docs by Speaking." Wired, 29 February. https://www.wired.com/2016/02/now-can-type-google-docs-speaking/
8. Munteanu, E., Guggiana, V., Darvishi, A., Schauer, H., Rauterberg, G. W. M., & Motavalli, M. (1995). Physical modelling of environmental sounds. In F. Pedrielli (Ed.), Proceedings of the 2nd international conference on acoustic and musical research, CIARM '95 (pp. 107-112). Universita di Ferrara.
9. Oliveira, J. Guerreiro, T. Nicolau, H. Jorge, J. Gonçalves, D. (2011). BrailleType: unleashing braille over touch screen mobile phones. IFIP Conference on Human-Computer Interaction, Springer, Berlin, Heidelberg, pp. 100-107.

10. Rajkumar, A., J. Lazar, J. B. Jordan, A. Darvishy, and H. Hutter. (2020). PDF Accessibility of Research Papers: What Tools are Needed for Assessment and Remediation?. In HICSS.
11. Sabharwal, N., Agrawal, A. (2020). Introduction to Google Dialogflow. In: Cognitive Virtual Assistants Using Google Dialogflow. Apress, Berkeley, CA.
12. World Health Organization (WHO). (2012). Global Data on Visual Impairments 2010. Geneva, World Health Organization.